\documentclass[10pt,letterpaper,twocolumn]{article} %% two column, final layout

\usepackage{ol2}
\usepackage{color}
\usepackage[draft,implicit=false]{hyperref}
\usepackage{amsmath}
\usepackage{amsfonts}
\usepackage{amssymb}
\usepackage{graphicx}
\usepackage{bm}

\begin{document}

\twocolumn[ %% activate for two-column option

\title{Spatial sub-Rayleigh imaging analysis via speckle laser illumination}

%% For REVTeX it is possible to automate superscript and e-mail callouts with the superscriptaddress option; see REVTeX4 documentation.

\author{Yunlong Wang, Feiran Wang, Ruifeng Liu, Dongxu Chen, Hong Gao, Pei Zhang,$^{\ast}$ and Fuli Li}

\address{
MOE Key Laboratory for Nonequilibrium Synthesis and Modulation of Condensed Matter, Department of Applied Physics,

Xi'an Jiaotong University, Xi'an 710049, People's Republic of China.\\
$^*$Corresponding author: zhangpei@mail.ustc.edu.cn
}

\begin{abstract}

It is commonly accepted that optical sub-Rayleigh imaging has potential application in many fields.
In this paper, by confining the divergence of the optical field as well as the size of the illumination source, we show that the first-order averaged intensity measurement via speckle laser illumination can make an actual breakthrough on the Rayleigh limit.
For high-order algorithm, It has been reported that the autocorrelation function can be utilized to achieve the sub-Rayleigh feature.
However, we find that this sub-Rayleigh feature for the high-order algorithm is limited only to binary objects and the image will be distorted when a gray object is placed. 
This property encourages us to find the physics behind the high-order correlation imaging algorithm.
We address these explanations in this manuscript and find that for different types of high-order algorithm, there is always a ``seat'' in the right place from the cross-correlation function.

\end{abstract}


] %% activate for two-column option
 
\noindent Optical imaging is an indispensable tool in practice.
The key advantages, especially for far-field, over other forms of imaging are the capability and compatibility of non-contact, minimally invasive observation and real-time feedback.
Unfortunately, due to the inherent diffraction barrier, it diffuses a point in the object plane into the Airy disk, which has larger size than the original point.
The image can never perfectly represent the real details and the resolution limit is defined as Rayleigh limit.

People need further details by means of the imaging techniques with fewer sacrifice or less physical constraints.
During the past decades, several techniques based on classical or quantum principle have been introduced to improve the resolution and they may fall into three classes as follows.
Changing the imaging mode by a lensless scheme is built toward aberration-free diffraction-limited image that avoid the limit of lens-based systems, such as ghost imaging\cite{PRL2005,PRA2007} and memory-effect imaging\cite{ME} via speckle correlation.
Changing the imaging units is to receive additional components of the object spectrum, notably, structured illumination microscopy (SIM)\cite{SIM} and super-lens imaging\cite{superlens}.
Exploiting post-selection algorithm is to recover the discarding part of the measurement data by extracting a nonclassical component from classical-state light which contains the imaging information, for instance, sparsity algorithm\cite{OE209,OL210,OL2014} and high-order correlation algorithm\cite{APL2008,OL2009}.
These three classes of techniques can achieve the sub-Rayleigh features by receiving the object information more efficiently without the need for getting over the affect of the propagation process.

To combine these three techniques in a hybrid arrangement, it is proposed in quantum laser radar scan technique\cite{PRL2010,OE2011} that designs an active imager namely a focused laser point projecting onto subportions of an object just like the pointillism and a receiver to localize the center of the illumination point.
By reducing the size of the illumination light as well as the point spread function (PSF), this scheme lends the above-mentioned effort in localizing the imaging point accurately so that the point-to-point perfect imaging is achieved.
Accordingly, we can also confine the the divergence of the optical field to make it possible, for instance, by utilizing the speckle-free laser imaging scheme\cite{NP,PNAS2015}. 
Based on this principle, we will suggest  how to design an approach of sub-Rayleigh imaging by controlling the optical field and the size of the illumination light effectively in the next.

For the high-order imaging, J. E. Oh \textit{et al} utilized the autocorrelation to reduce the PSF as a fact of $ 1/\sqrt{2} $ to extract the object information covered by the Airy disk during the imaging process\cite{OL2013}.
However, some major issues such as the affect of the object function squared, the physical description behind the super-resolution algorithm are still unclear.
In this manuscript, we will provide a further theoretical explanation for the spatial imaging, high-order algorithm and make a full understanding through the experiment demonstration.

We start by reviewing some basics of the propagation process.
Suppose that a series of $z$-propagating, monochromatic and random laser speckles with the size of the source $D_s$ emits from plane $z=0$.
These speckles transilluminate an object $T(\bm{\rho}_0)$ at $\bm{\rho}_0$ planar plane $d_s$ away from the source.
The light passes through the object is then collected by a diffraction-limited circular imaging lens of diameter $D$ located at $d_0$ in front of the object.
This lens casts an image of the object at a far-field distance (the Fraunhofer zone according to the incoherent light) $d_i$ beyond the imaging lens and limits the imaging precision by its numerical aperture ($N\!A$).
The schematic of the imaging protocol is shown in Fig. 1.
To exhibit the sub-Rayleigh capability of this scheme, we shall assume the light on the source plane as fully spatially incoherent namely infinitesimal transverse coherence length during the propagation.
According to the Fresnel diffraction formula and statistical law of optical fields, by neglecting the propagation coefficient and phase delay, the first-order intensity distribution guided to the image plane is calculated to be

\begin{align}
I(\bm{\rho},\bm{\rho}')\propto &\int_{T}\int_{T'}d\bm{\rho}_{0}d\bm{\rho}_{0}'T^2(\bm{\rho}_{0})\nonumber \\
&\times{somb(\frac{\pi{D_s}}{m\lambda{d_s}}|\bm{\rho}_{0}-\bm{\rho}_{0}'|) \label{Eq2}} \nonumber \\
&\times{somb(\frac{\pi{D}}{m\lambda{d_0}}|\bm{\rho}_{0}-\frac{\bm{\rho}}{M}|)} \nonumber \\
&\times{somb(\frac{\pi{D}}{m\lambda{d_0}}|\bm{\rho}_{0}'-\frac{\bm{\rho}'}{M}|)},
\end{align}
where $M=d_i/d_0$ represents the imaging magnification, $somb(x)\equiv2J_1(x)/x$ is the Sombrero function which can be regarded as the PSF and $ J_1(x) $ is the first Bessel function. Here we add a variable $m$ discreetly on Eq. (1) to characterize the divergence of the illumination.
This parameter makes the quantification of the optical propagation and it is related to the illumination source, lens and so on. When $m=0$, the Sombrero function can be approximated as the delta function and it means the parallel beam.
$m=1$ represents the normal Fresnel propagation.
When $0<m<1$, it shows the compression of the Fresnel propagation such as the laser beam as well as $m>1$ indicates a faster expansion than the Fresnel case, for instance, the combination of the source and a concave lens.

Conceptually, Eq. (\ref{Eq2}) clearly shows that the resolution of the first-order limits by three elements: the divergence $m$, the Sombrero function of the light source and lens-imaging\cite{OL2009}.
As the transverse coherence length ($lc$) of the pseudo-thermal source $lc=m\lambda{d_s/D_s}$ reduced or $N\!A=D/d_0$ expanded, the resolution of the imaging system can be improved and will be better than each subitem exists on its own.
It implies that under the same conditions, this system has smaller spatial extent compared to the conventional imaging with lens-imaging Sombrero function only or the traditional ghost imaging system by a lensless scheme.
On the other hand, the product cautions us each subitem should not be neglected easily, for instance, the image will be blurry since a smaller $N\!A$ brings the speckle diffraction. 

Then we can obtain the second-order fluctuation cross-correlation function 
\begin{align}
&\Delta{G^{(2)}(\bm{\rho}_A,\bm{\rho}_B)}=<\Delta I(\bm{\rho}_A)\Delta I(\bm{\rho}_B)> \nonumber \\
&\qquad\qquad\qquad \propto\int_{T_A}\int_{T_B}d\bm{\rho}_{A_0}d\bm{\rho}_{B_0}T^2(\bm{\rho}_{A_0})T^2(\bm{\rho}_{B_0})\label{Eq3} \nonumber \\
&\qquad\qquad\qquad\quad \times{somb^2(\frac{\pi{D_s}}{m\lambda{d_s}}|\bm{\rho}_{B_0}-\bm{\rho}_{A_0}|)}\nonumber \\
&\qquad\qquad\qquad\quad \times{somb^2(\frac{\pi{D}}{m\lambda{d_0}}|\bm{\rho}_{A_0}-\frac{\bm{\rho}_{A}}{M}|)}\nonumber \\
&\qquad\qquad\qquad\quad \times{somb^2(\frac{\pi{D}}{m\lambda{d_0}}|\bm{\rho}_{B_0}-\frac{\bm{\rho}_{B}}{M}|)}.
\end{align}
When $\bm{\rho}_{A}=\bm{\rho}_{B}$, the autocorrelation function\cite{OL2013} is given by  
\begin{align}
\Delta G^{(2)}(\bm{\rho},\bm{\rho}')\propto &\int_{T}\int_{T'}d\bm{\rho}_{0}d\bm{\rho}_{0}'T^4(\bm{\rho}_{0})\nonumber \\
&\times{somb^2(\frac{\pi{D_s}}{m\lambda{d_s}}|\bm{\rho}_{0}-\bm{\rho}_{0}'|)\label{Eq4}} \nonumber \\
&\times{somb^2(\frac{\pi{D}}{m\lambda{d_0}}|\bm{\rho}_{0}-\frac{\bm{\rho}}{M}|)}\nonumber \\
&\times{somb^2(\frac{\pi{D}}{m\lambda{d_0}}|\bm{\rho}_{0}'-\frac{\bm{\rho}'}{M}|)},
\end{align}
here the autocorrelation function can be regarded as the modular square of the first-order.
It means a $1/\sqrt{2}$ value further decrease by squeezing the full width at half maximum (FWHM) of the Sombrero function.\\

\begin{figure}[htb]
\centerline{\includegraphics[width=8.0 cm]{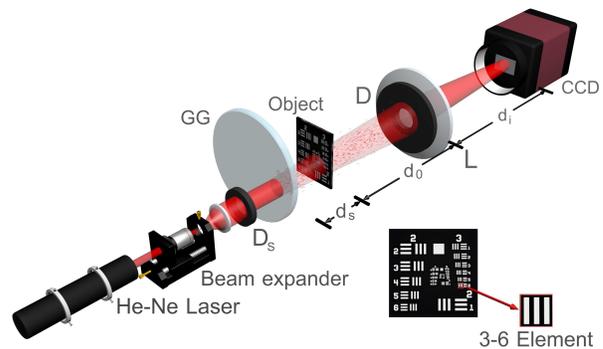}}
\caption{Schematic of the sub-Rayleigh imaging system.}
\end{figure}

We demonstrate the concept of sub-Rayleigh imaging with the setup shown in Fig. 1.
The speckle laser is simulated by projecting a collimated (modified by the beam expander and the aperture $D_s$) He-Ne laser beam $\lambda=632.8$ nm onto a slowly rotating ground glass (GG). 
The object to be imaged in transmission is the triple-slit (group 3, element 6 of the USAF resolution gauge composed of opaque and clear stripes $35\ \mu$m wide), which is imaged through an $f=150$ mm diffraction-limited imaging lens (L) setting in a tunable aperture to adjust $N\!A$. The detector is the Charge Coupled Device (CCD) with $1392\times1040$ array of $6.45\times6.45\ \mu$m$^2$ pixels. 
The magnification factor is fixed on $M=1$ ($d_0=d_i=300$\ mm) and the Rayleigh limit of the lens-imaging system is defined as $ \delta{x}\equiv{1.22\lambda{M}/{N\!A}} $.
With this setup, we can measure three different types of image in turn: the conventional image, the first-order average intensity and the second-order fluctuation correlation as defined in Eqs. (\ref{Eq2}) $\sim$ (\ref{Eq4}). 

Typical experimental results are presented in Fig. 2.
Originally we use the scatting light (the beam expander without the latter collimated lens and here $m=1$) illumination and set the diameter of the imaging lens L as $D=4$\ mm namely the Rayleigh limit $ \delta{x}=57.9\ \mu$m, larger than the triple-slit separation $35\ \mu$m.
Hence, a blurry conventional image is obtained in Fig. 2(a).
After the light collimated as laser beam ($m=0$), the image in Fig. 2(b) become more clear.
Then when using the speckle laser illumination ($0<m<1$, since the speckle has less FWHM compared with the scatting light after the same distance) with $lc=240 \ \mu$m, we can get a completely resolved image in Fig. 2(c).
Here the first step is to clarify the affect of the divergence and size of illuminated light.
During the second step about zooming out the diameter $D$ of L to 3.7\ mm and 3.2\ mm, a relatively resolved and blurry image are exhibited in Fig. 2(d)$\&$(e), respectively.
Keeping $D=3.2$\ mm constantly, the first two steps can be repeated in the next, the image can be resolved clearly again with the decrease of $lc$ as $lc=58\ \mu$m in Fig. 2(f).
The barely unresolved image is reproduced in Fig. 2(g) by zooming out as $D=2.7$\ mm and the resolution can be improved further shown in Fig. 2(h) with the decrease of $lc$ as $lc=13\ \mu$m. The whole process can be repeated inversely.
As equal to the analysis in Eq. (\ref{Eq2}), we make a full demonstration on the affect of the divergence of optical field as well as the size of illumination light $lc$ and the imaging lens $N\!A$.
In addition, based on the feature of the divergence, the experimental results show that the first-order measurement can surpass the Rayleigh limit drastically.

\begin{figure}[htb]
\centerline{\includegraphics[width=8.0 cm]{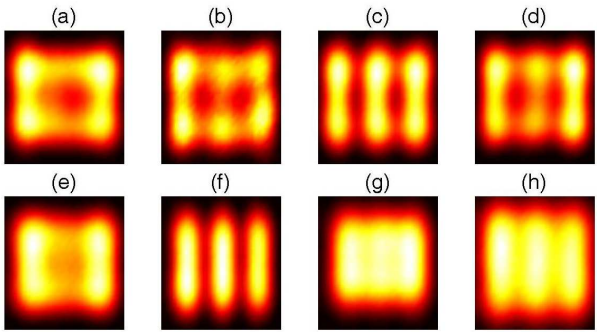}}
\caption{The first-order averaged intensity of the triple-slit measured by CCD. First step, illuminated by scatting light ($m=1$), collimated light ($m=0$) and speckle laser ($0<m<1$) with $lc=240\ \mu$m and the same diameter $D=4$\ mm of the imaging lens shown as (a)$\sim$(c), respectively. Second step, decrease the diameter shown as (d) $D=3.7$\ mm and (e) $D=3.2$\ mm. Last step, keep $D=3.2$\ mm with the decrease of $lc=58\ \mu$m shown in (e); keep $lc=58\ \mu$m and zoom out $D=2.7$\ mm in (g); keep $D=2.7$\ mm but  decrease $lc=13\ \mu$m in (h). %The image shown in (c)$\sim$(h) are measured averaged over $N=1000$ CCD frames with $1\ ms$ exposure time.
}
\end{figure}

In the following, we turn to the second-order imaging.
Analogous to the analysis of Ref.\cite{OL2013}, all $N$ frames are averaged pixel by pixel.
The average value is then subtracted from each frame, pixel by pixel, leaving only fluctuation terms.
Assuming the triple-slit as a one dimensional object, we evaluate the integral along the vertical coordinates $y_A$ and $y_B$ from each frame, and then build the fluctuation frame as labels of the horizontal coordinates $x_A$ and $x_B$. These results are summed over for all $N$ frames to give the cross-correlation outcome which are shown in Fig. 3(a)$\&$(c), where $x_A$ and $x_B$ are the horizontal coordinates on CCD plane.
Here we use color to represent the value of the normalized second-order correlation function.
These figures show an second-order interference pattern of the triple-slit by $\Delta{G^{(2)}(x_A,x_B)}$  \cite{SR2014}. %which can be treated as the projection of the 5-dimension cross-correlation function $\Delta{G^{(2)}(\bm{\rho}_A,\bm{\rho}_B)}$ \cite{SR2014}.

We try to use two different triple-slits, one is binary object with $1:1:1$ transparence degree and the other is gray object with $1:0.4:0.1$ transparence degree, to deeply understand the similarity and difference between high-order imaging and first order imaging. The second-order interference patterns of them are shown in Fig. 3(a)$\&$(c), respectively.
When we integrate along the vertical coordinate $x_B$, it provides the cross-section of the first-order measurement drawn in black lines in Fig. 3(b)$\&$(d), which are equivalent to the actual first-order imaging.
Then, under the condition of $x_A=x_B$, we extract the diagonal data from Fig. 3(a)$\&$(c) shown as red lines which indicates the autocorrelation function in Eq. (\ref{Eq4}).
Compared with the first-order lines, the autocorrelation demonstrates a resolution enhancement by a factor of $\sqrt{2}$, to say, due to the squeeze of the Sombrero function as expected, or to say, the length relationship between the hypotenuse and its side. However, we find that 
the reconstruction of the autocorrelation pattern is NOT the normal image because the term $T^4$ from Eq. (\ref{Eq4}) will result in an imaging distortion shown as Fig. 3(c)$\&$(d).
What is more, high-order correlation brings degradation in signal-to-noise \cite{PNAS2009,OL2010}.
Obviously, the autocorrelation inversion (the square root) returns to the original feature of all aspects in the first-order. So we can conclude that the high-order correlation imaging only benefits to the binary object.
%On the other hand, the autocorrelation effort also can be regarded as the contrast enhancement \cite{PNAS2009,APL2014,OL2010}. 

\begin{figure}[htb]
\centerline{\includegraphics[width=8.0 cm]{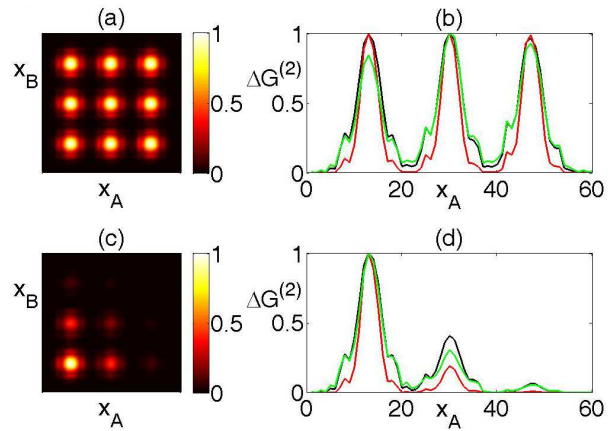}}
\caption{The second-order cross-correlation image of the triple slits. On the first column, (a)$\&$(c) show the binary ($1:1:1$) and gray ($1:0.4:0.1$) level object illumination, respectively. 
On the second column, (b)$\&$(d) are the cross-section patterns corresponding to the former. The lines drawn in black is treated as the first-order cross-section of the normal image while red lines are the diagonal data extracted from Fig. 3(a)$\&$(c) indicating the autocorrelation function. The green lines are for the PNFC algorithm results.}
\end{figure}

The last point we need to figure out is the physics behind the high-order algorithm. Here we need to classify the above cases into two groups according to the position of the object.
The object put behind the beam-splitter is the first group, for instance, ghost imaging with thermal\cite{PRL2005,PRA2007} and entangled source\cite{es}.
In this case, we can get the image from the correlation value $<\Delta{I_{test}}\Delta{I_{ref.}}>$, where $\Delta{I_{test}}(\Delta{I_{ref.}})$ represents the fluctuation of test (reference) plane with (without) the object.
There is a critical result that the imaging point is the enhancement point of the correlation value which can be regarded as the physics behind the correlation function between the different coordinates.
After the integral measurement of the test plane which can be regarded as the projection to the reference plane, we have the object information transferred and finally get the image.
For the second group, the object is put before the beam-splitter. In this case, the optical field of each plane is exactly the same. For the N-photon sub-wavelength detection\cite{SR2014} as well as the autocorrelation image\cite{OL2013}, the correlation value is no longer the image of the real object but the object information squared which affects the image quality namely the resolution, contrast and signal-to-noise.
As we know, the high-order correlation function is with multiple variables, and to get the image, we generally should project the correlation function to some particular planes\cite{SR2014}. The autocorrelation image chooses a relationship $\bm{\rho}=\bm{\rho}'$ as the projection plane, brings out the resolution enhancement of $somb^4(\frac{\pi{D}}{\lambda{d_0}}|\bm{\rho}_{0}-\frac{\bm{\rho}}{M}|)$\cite{OL2013} and the problem of image distortion (as we show in our experiment with gray objects) simultaneously.
The PNFC algorithm \cite{pnfc,arxiv} is another projection to the coordinates of one arm just like the ghost imaging. It is a recovery of the real object with high contrast but no resolution enhancement since no other physical principle introduced in.
In conclusion, we can study the image properties by tracing back to the corresponding correlation function.

In conclusion, based on the speckle laser illumination, we have proposed a scheme of sub-Rayleigh imaging via the first-order and second-order measurement.
By this scheme, the resolution improvement can be obtained by the decreasing the divergence of optical field as well as the size of illumination light. %and $ T^2 $ from Eq. (\ref{Eq1}) means this improvement aiming at the normal image.   
In our experiment, we find that for binary object, it can effectively prevent the image distortion and degradation in the signal-to-noise caused by the high-order imaging and is feasible for the resolution enhancement of the conventional imaging or the traditional ghost imaging system.
However, for gray object, the distortion of the autocorrelation imaging is occurred and the resolution improvement is not aiming at normal imaging because of $ T^4 \neq T^2$ anymore.
It will move toward the same condition of the first order in resolution by the square root.
%Moreover, the  features can be traced back to the mapping from the cross-correlation function.
%J. E. Oh et al. reported a clear enhancement in resolution by reducing the size of the speckle illumination $lc$ but this effect can be attributed to the first order, not the autocorrelation function. 
In addition, it is a useful tool to study high-order algorithm properties from the corresponding cross-correlation function and we can also expand this approach to the optical mode discrimination as well as the spectrum analysis.

This work is supported by the Fundamental Research Funds for the Central Universities, Program for Key Science and Technology Innovative Research Team of Shaanxi Province (No. 2013KCT-05), and the National Natural Science Foundation of China ( Grant Nos. 11374008, 11374238 and 11374239).

\pagebreak

\section*{Informational Fifth Page}


\begin{thebibliography}{99}
% Ghost imaging
\bibitem{PRL2005} F. Ferri, D. Magatti, A. Gatti, M. Bache, E. Brambilla, and L. A. Lugiato, ``High-resolution ghost image and ghost diffraction experiments with thermal light,'' Phys. Rev. Lett. \textbf{94,} 183602 (2005).

\bibitem{PRA2007} M. H. Zhang, Q. Wei, X. Shen, Y. F. Liu, H. L. Liu, J. Cheng, and S. S. Han, ``Lensless Fourier-transform ghost imaging with classical incoherent light,'' Phys. Rev. A \textbf{75,} 021803(R) (2007).
% momery effect
\bibitem{ME} O. Katz, P. Heidmann, M. Fink, and S. Gigan, ``Non-invasive single-shot imaging through scattering layers and around corners via speckle correlations,'' Nat. photonics \textbf{8,} 784 (2014).
% SIM
\bibitem{SIM} D. Dan, M. Lei, B. L. Yao, W. Wang, M. Winterhalder, A. Zumbusch, Y. J. Qi, L. Xia, S. H. Yan, Y. L. Yang, P. Gao, T. Ye, and W. Zhao, ``DMD-based LED-illumination super-resolution and optical sectioning microscopy,'' Scientific Reports \textbf{3,} 01116 (2013).
% superlens
\bibitem{superlens} E. G. van Putten, D. Akbulut, J. Bertolotti, W. L. Vos, A. Lagendijk, and A. P. Mosk, ``Scattering lens resolves sub-100 nm structures with visible light,'' Phys. Rev. Lett. \textbf{106,} 193905 (2011).
%+sparsity constraints 
\bibitem{OE209} S. Gazit, A. Szameit, Y. C. Eldar and M. Segev, ``Super-resolution and reconstruction of sparse sub-wavelength images,'' Opt. Express \textbf{17,} 23920 (2009).

\bibitem{OL210} Y. Shechtman, S. Gazit, A. Szameit, Y. C. Eldar and M. Segev, ``Super-resolution and reconstruction of sparse images carried by incoherent light,'' Opt. Lett. \textbf{35,} 1148 (2010).

\bibitem{OL2014} A. D. Rodríguez, P. Clemente, E. Irles, E. Tajahuerce and J. Lancis, ``Resolution analysis in computational imaging with patterned illumination and bucket detection,'' Opt. Lett. \textbf{39,} 3888 (2014).
%+high-order
\bibitem{APL2008} D. Z. Cao, J. Xiong, S. H. Zhang, L. F. Lin, L. Gao, and K. G. Wang, ``Enhancing visibility and resolution in Nth-order intensity correlation of thermal light,'' Appl. Phys. Lett. \textbf{92,} 201102 (2008).

\bibitem{OL2009} P. L. Zhang, W. L. Gong, X. Shen, D. J. Huang, and S. S. Han, ``Improving resolution by the second-order correlation of light fields,'' Opt. Lett. \textbf{34,} 1222 (2009).

% laser lidar
\bibitem{PRL2010} F. Guerrieri, L. Maccone, F. N. Wong, J. H. Shapiro, S. Tisa, and F. Zappa, ``Sub-Rayleigh imaging via N-photon detection,'' Phys. Rev. Lett. \textbf{105,} 163602 (2010).

\bibitem{OE2011} S. Mouradian, F. N. Wong, and J. H. Shapiro, ``Achieving sub-Rayleigh resolution via thresholding,'' Opt. Express \textbf{19,} 5480 (2011).

% Caohui
\bibitem{NP} B. Redding, M. A. Choma, and H Cao, ``Speckle-free laser imaging using random laser illumination,'' Nat. Photonics \textbf{6,} 355 (2012).

\bibitem{PNAS2015} B. Redding, A. Cerjan, X. Huang, M. L. Lee, A. D. Stone, M. A. Choma, and H. Cao, ``Low spatial coherence electrically pumped semiconductor laser for speckle-free full-field imaging,'' Proc. Natl. Acad. Sci. USA \textbf{112,} 1304 (2015).

%Kim
\bibitem{OL2013} J. E. Oh, Y. W. Cho, G. Scarcelli, and Y. H. Kim, ``Sub-Rayleigh imaging via speckle illumination,'' Opt. Lett. \textbf{38,} 682 (2013).
%visibility
\bibitem{PNAS2009} T. Dertinger, R. Colyer, G. Iyer, S. Weiss and J. Enderlein, ``Fast, background-free, 3D super-resolution optical fluctuation imaging (SOFI),'' Proc. Natl. Acad. Sci. USA \textbf{106,} 22287 (2009).

\bibitem{OL2010} X. H. Chen, I. N. Agafonov, K. H. Luo, Q. Liu, R. Xian, M. V. Chekhova, and L. A. Wu, ``High-visibility, high-order lensless ghost imaging with thermal light,'' Opt. Lett. \textbf{35,} 1166 (2010).


%% mutiphotons
%%PNFC
%\bibitem{arxiv} J. N. Sprigg, T. Peng, Y. H. Shih, ``Nonclassical imaging via photon number fluctuation correlation,'' arXiv \textbf{1409,} 2134 (2014).
%
%\bibitem{pnfc} H. Chen, T. Peng, and Y. H. Shih, ``100\% correlation of chaotic thermal light,'' Phys. Rev. A \textbf{88,} 023808 (2013).
%
%\bibitem{SR2014} R. F. Liu, P. Zhang, Y. Zhou, H. Gao, and F. L. Li, ``Super sub-wavelength patterns in photon coincidence detection,'' Sci. Rep. \textbf{4,} 4068 (2014).
%
%\bibitem{APL2014} Y. H. Zhai, F. E. Becerra, J. Y. Fan, and A. Migdall, ``Direct measurement of sub-wavelength interference using thermal light and photon-number-resolved detection,'' Appl. Phys. Lett. \textbf{105,} 101104 (2014).

% entangled source
\bibitem{es} G. B. Lemos, V. Borish, G. D. Cole, S. Ramelow, R. Lapkiewicz and A. Zeilinger, ``Quantum imaging with undetected photons,'' Nat. \textbf{512,} 409 (2014).
% mutiphotons
\bibitem{SR2014} R. F. Liu, P. Zhang, Y. Zhou, H. Gao, and F. L. Li, ``Super sub-wavelength patterns in photon coincidence detection,'' Sci. Rep. \textbf{4,} 4068 (2014).
%PNFC
\bibitem{arxiv} J. N. Sprigg, T. Peng, Y. H. Shih, ``Nonclassical imaging via photon number fluctuation correlation,'' arXiv \textbf{1409,} 2134 (2014).
\bibitem{pnfc} H. Chen, T. Peng, and Y. H. Shih, ``100\% correlation of chaotic thermal light,'' Phys. Rev. A \textbf{88,} 023808 (2013).

\end{thebibliography}
\end{document}